\documentclass[twocolumn,
aps,nofootinbib,showpacs,showkeys,preprint
tightenlines,preprintnumbers,
,superscriptaddress
] {revtex4}

\usepackage{epsf,epsfig,subfigure,graphicx,amsmath,amssymb}
\usepackage{color}
\newcommand{\dis}[1]{\begin{equation}\begin{split}#1\end{split}\end{equation}}
\newcommand{\OVER}[1]{\,\overline{\hskip -0.5mm #1}}

\newcommand{\FSI}[1]{\phi^{\rm FSI}_{#1}}

\newcommand{\ie}{{\it i.e.}\ }
\newcommand{\etal}{{\it et al.} }
\newcommand{\Qem}{Q_{\rm em}}

\newcommand{\gev}{\,{\rm GeV}}

\begin{document}
\title{\bf Final state interaction phases using data from CP asymmetries}
\author{Jihn E.  Kim}
\address{Department of Physics, Kyung Hee University, 26 Gyungheedaero, Dongdaemun-Gu,
Seoul 130-701, Korea,
}
\author{Doh Young Mo}
\address{Department of Physics, Seoul National University, 1 Gwanakro,  Gwanak-Gu,
Seoul 151-747, Korea
}
\author{Soonkeon Nam$^1$}

\begin{abstract}
Evidence of physics beyond the Standard Model (SM) of particle physics might be revealed through CP violation. Experimental results from BaBar, Belle, and most recently from LHCb are very interesting in this respect. It is important to calculate various CP violation parameters within the SM.  In this Letter, the final state interaction (FSI) phases with the CKM matrix  are calculated using the recent experimental results from LHCb on CP violation in $\OVER{B}^{\,0}$ and $\OVER{B}^{\,0}_{s}$ meson decays. To obtain the allowed regions of the FSI phases, a simplified form of the Jarlskog determinant of the CKM matrix is used and the CP phase $\delta$ close to the maximal value of $\pi/2$ is found. Then, the asymmetry data on $\OVER{B}^{\,0}$ and $\OVER{B}^{\,0}_{s}$ from LHCb is used to explore graphically the allowed regions of the FSI phases.

\end{abstract}

\maketitle

{\it Introduction} --
With the discovery of the Higgs all the SM particles have now been detected. Notwithstanding this huge success, one of the outstanding problems in particle physics is a full explanation of the Cabibbo-Kobayashi-Maskawa (CKM) matrix \cite{Cabibbo63, KM73} which encodes the quark mixing.  The CKM matrix can be parametrized with three real (CP- conserving) angles $\theta_i (i=1,2,3)$ and one imaginary (CP-violating) phase \cite{CPphase} $\delta$.  In addition, in the presence of hadronic CP-violating processes, we introduce final state interaction (FSI) phases $\FSI{j}$.  Precision measurements of these parameters and comparison with the SM predictions are sensitive probes of new physics beyond the SM.

There are several ways to parametrize the CKM matrix \cite{Maiani76, Wolfenstein83, ChauPRL84}.  Among them, the Wolfenstein parametrization  \cite{Wolfenstein83} has been widely used because of its hierarchical representation.  However, the physics of weak CP violation is represented by the Jarlskog determinant $J$ \cite{Jarlskog85, Jarlskog2}. $J$ encodes the weak CP violation  in an invariant manner. In this respect, the recent direct observations [9] of CP violation in B0 and B0s decays into $K + \pi$ modes are very interesting.

In this Letter, we introduce a new form of the Jarlskog determinant (which we will call the KS form), and find that we can determine $\delta$ with high precision. Next, by explicitly including the FSI phases in the formulation, we can compare our calculation of the decay rates of B mesons with the LHCb data.  We find for the first time the allowed values of the FSI phases by assuming a reasonable parameter range. Until now, the SM confirmation of the LHCb data was consistent with the SM prediction $\Delta\approx 0$  \cite{LHCbAsymm}, the Lipkin’s variable\cite{Lipkin05}.  However, this variable $\Delta$ is defined such that the unknown FSI phases,  $\FSI{1/2,3/2}$, etc., are ignored.

{\it New form of Jarlskog determinant} --
The Jarlskog determinant was originally expressed as the imaginary part of a product of two elements of $V$ and two elements of $V^*$. Even if $V$ is complex, the determinant of $V$ can be real. If  the determinant is complex, one can make it real by multiplying a common phase to all the $\Qem=2/3$ or to all the $\Qem=-1/3$ quarks. Even if the determinant is real, there can be CP violation phenomena because each of the six terms in the determinant is complex. The magnitude of the imaginary part of each of the six terms is the same. It is the Jarlskog determinant, for example $J=|{\rm Im\,}V_{31}V_{22}V_{13}|$ \cite{KimSeoJarl12}. We will introduce just one phase $\delta$ in the CKM matrix and parametrize $V_{ij}$ such that the first row is real. Because $V_{22}$ is close to 1, the phase in $V_{31}$ is interpreted as the weak CP phase.

The simple form of the Jarlskog determinant $J=|{\rm Im \,}V_{31}V_{22}V_{13}|$ has not been derived for almost three decades \cite{BigiSandaBk, BrancoBk, PichARMP, NirKirkby}.
The Jarlskog determinant is twice the area of the Jarlskog triangle. If two quark masses are degenerate, there is a redefinition freedom of the degenerate quarks and hence the CP phase becomes a physically unobservable one. The Jarlskog determinant contains the phase $\delta$ of the CKM matrix. Since observation of the CP violation is an interference phenomena, if the CKM phase $\delta$ is 0 or $\pi$ then there is no interference and hence no CP violation.

For the real and hence the unit determinant, ${\rm Det.}\,V =1=
 V_{11}V_{22}V_{33} -V_{11}V_{23}V_{32} +V_{12}V_{23}V_{31}
 -V_{12}V_{21}V_{33} +V_{13}V_{21}V_{32} -V_{13}V_{22}V_{31}$.
Multiply by $V_{13}^*V_{22}^*V_{31}^*$ on both sides. Then, we obtain
\dis{V_{13}^* &V_{22}^*V_{31}^* =|V_{22}|^2V_{11}V_{33}V_{13}^*V_{31}^*-V_{11}
 V_{23}V_{32}V_{13}^*V_{31}^*V_{22}^*\\
 &+|V_{31}|^2V_{12}V_{23}V_{13}^*V_{22}^* -V_{12}V_{21}V_{33}V_{13}^*V_{31}^*V_{22}^*\\
 &+|V_{13}|^2V_{21}V_{32}V_{31}^*V_{22}^*-|V_{13}V_{22}V_{31}|^2.\nonumber
 }
Using the unitarity of $V$ as discussed in \cite{KimSeoJarl12}, the equation can be rewritten as
 \dis{
  &V_{13}^*V_{22}^*V_{31}^* =(1-|V_{21}|^2)V_{11}V_{33}V_{13}^*V_{31}^* \\
 &~+V_{11}V_{23} V_{13}^*V_{21}^*|V_{31}|^2 +(1-|V_{11}|^2)V_{12}V_{23}V_{13}^*V_{22}^*\\
 &~+|V_{13}|^2(V_{12}V_{21}V_{11}^*V_{22}^*+V_{21}V_{32}V_{31}^*V_{22}^*)\\
 &~-|V_{13}V_{22}V_{31}|^2.
 \label{eq:detmult3}
 }
Let the imaginary part of $V_{11}V_{33}V_{13}^*V_{31}^*$ be $J$. Now, using the unitarity relations again, we can express the imaginary part of the RHS of Eq. (\ref{eq:detmult3}) as
$[(1-|V_{21}|^2)-|V_{31}|^2+(1-|V_{11}|^2)]J=J$ \cite{KimSeoJarl12}.
Therefore, the imaginary part of  $V_{13}^*V_{22}^*V_{31}^*$ (the LHS of Eq. (\ref{eq:detmult3})) is $J$.  It is {\it the imaginary part of any one element among the six components of determinant of $V$}, for example  $J=|{\rm Im}\,V_{13} V_{22} V_{31} |$, which will be called the Kim-Seo(KS) form  \cite{KimSeoJarl12}. This simplifies the method to scrutinize the weak CP violation effects just looking at the CKM matrix elements. Making the elements of the first row real, the phase of $V_{31}$ is an invariant phase. A CKM matrix with a real determinant is chosen as
\dis{
 \left(\begin{array}{ccc} c_1,&s_1c_3,&s_1s_3 \\ [0.2em]
 -c_2s_1,&e^{-i\delta}s_2s_3 +c_1c_2c_3,&-e^{-i\delta} s_2c_3+c_1c_2s_3\\[0.2em]
-e^{i\delta} s_1s_2,&-c_2s_3 +c_1s_2c_3 e^{i\delta},& c_2c_3 +c_1s_2s_3 e^{i\delta}
\end{array}\right) \label{eq:KSexact}
}
where $c_i=\cos\theta_i$ and $s_i=\sin\theta_i$. We will call the phase $\delta$ appearing in $V_{31}$ the {\it Jarlskog invariant phase} since it is the physical phase describing the strength of the weak CP violation.

\begin{figure}[!t]
  \begin{center}
  \begin{tabular}{c}
   \includegraphics[width=0.17\textwidth]{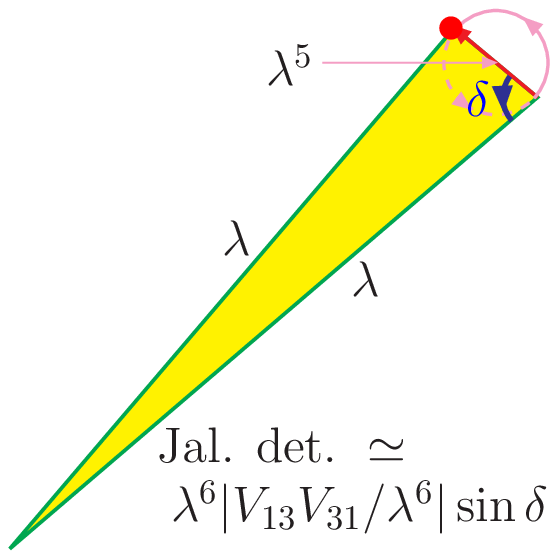} \\ \hskip 4.5cm (a)\\
    \includegraphics[width=0.25\textwidth]{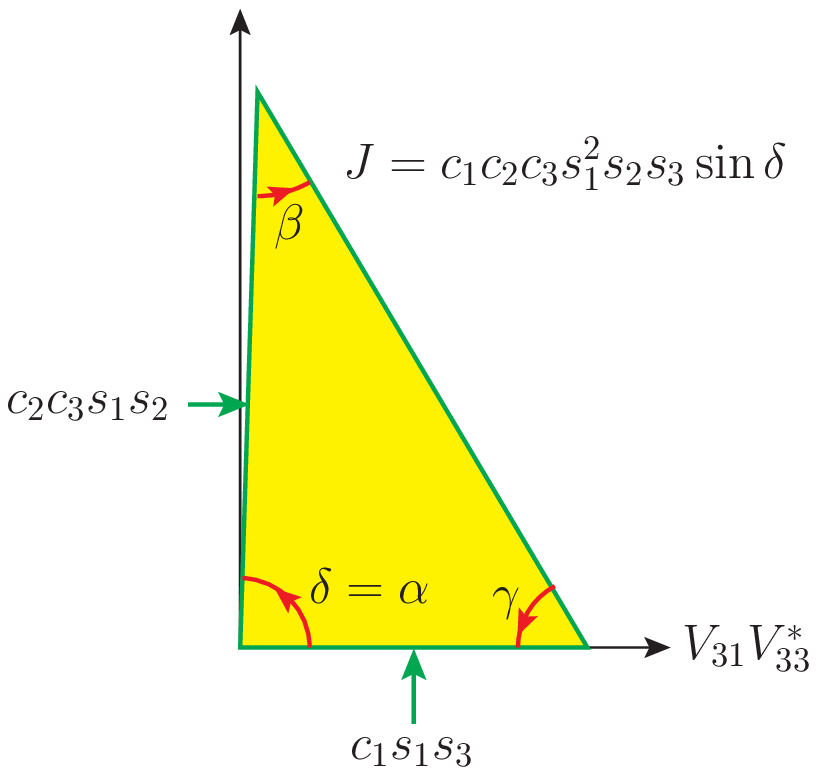} \\ \hskip 4.5cm  (b)
  \end{tabular}
  \end{center}
 \caption{The Jarlskog triangles. (a) The Jarlskog triangle with the first and second columns: this triangle has two long sides of $O(\lambda)$. Rotating the $O(\lambda^5)$ side (the red arrow), the CP phase $\delta$ changes. (b) The Jarlskog triangle with the first and third columns: the green base line is real, coming from $V_{11}V_{13}^*$ of Eq. (\ref{eq:KSexact}). The magnitude of the almost vertical green line comes from the coefficient of the factor $e^{-i\delta}$ of $V_{33}V_{31}^*$ of Eq. (\ref{eq:KSexact}).
  }
\label{fig:JTriLong}
\end{figure}

The physical magnitude of the weak CP violation is given by the area of the Jarlskog triangle, half of $J$. For any Jarlskog triangle, the area is the same. With the $\lambda= \sin\theta_1\equiv \sin\theta_C$ expansion, the area of the Jarlskog triangle is of order $\lambda^6$. In Fig. \ref{fig:JTriLong}\,(a), we show the triangle with two long sides (for the case of the first and second columns) of order $\lambda$. Rotating the $O(\lambda^5)$ side (the red arrow), the CP phase $\delta$ and also the area change. The magnitude of the Jarlskog determinant is $J\simeq\lambda^6|V_{13}V_{31}/\lambda^6|\sin\delta$. From  Fig. \ref{fig:JTriLong}\,(a), we notice that the area is maximum for  $\delta\simeq\frac{\pi}{2}$, and the maximality $\delta\simeq \frac\pi{2}$ is a physical statement. As $\delta$ is rotated in  Fig. \ref{fig:JTriLong}\,(a), the Jarlskog triangle of Fig.  \ref{fig:JTriLong}\,(b) also rotates and its area becomes maximal when $\delta\simeq \pi/2$. In any other parametrization of the CKM matrix, the same conclusion on maximality would result. The maximal CP phase was anticipated in \cite{GeorShin85} and can be modeled as shown in \cite{KimPLB11}.

The maximal CP violation is when $J= |\frac14 (c_1-c_1^3) \sin2\theta_2 \sin2\theta_3  \sin\delta| $ takes the maximal value. This occurs, in case of the 1st quadrant angles, for $\theta_1=\cos^{-1}(1/\sqrt3), \theta_2=45^0, \theta_3=45^0$, and $\delta=90^0$, and $J_{\rm max}=1/6\sqrt3$. The experimental value is $\approx\lambda^6\kappa_b\kappa_t\sin\delta\simeq 2.96^{+0.20}_{-0.16}\,\times{10^{-5}}$  \cite{KimSeo11, PDG12} which is about $3.1\times 10^{-4}$ fraction of $J_{\rm max}$. Even though the CP phase is maximal, the reduction of $J$ from $J_{\rm max}$ is due to the smallness of $\lambda$, \ie due to the  almost diagonal aspect of the CKM matrix \cite{PMNScomm}.


{\it The $\OVER{B}_{[s]}^{\,0}$ decay asymmetries} --
In the calculation of the direct CP violation, there occurs the strong FSI phases $\FSI{i}$.  Because of this strong phases, so far it was difficult to obtain a bound on $\delta$ from the measurements on the direct CP violation. See, for example, Ref. \cite{BaekS09}. Since the Jarlskog invariant $\delta$ has been measured rather accurately now, we can use it to predict the direct CP violation measurements on  $\FSI{i}$ or on the asymmetry factors with the full allowed range of $\FSI{i}$. For the $B_{d,s}^{0}$ decays \cite{Foot1}, the asymmetries have been measured \cite{LHCbAsymm},
\dis{
&A^{\rm CP,\, LHCb}_{\OVER{B}_d^{\,0}\to \pi^-K^+}= -0.080\pm 0.007({\rm stat})\pm 0.003({\rm syst}),\\
&A^{\rm CP,\, LHCb}_{\OVER{B}_s^{\,0}\to \pi^+K^-}= + 0.27\pm 0.04({\rm stat})\pm 0.01({\rm syst}),
\label{eq:expValues}
}
where
\dis{
&A^{\rm CP}_{\OVER{B}_d^{\,0}\to \pi^+K^-}=\frac{{\rm Br}(\OVER{B}_d^{\,0}\to \pi^+K^-) -{\rm Br}(B_d^0\to \pi^-K^+)}{{\rm Br}(\OVER{B}_d^{\,0}\to \pi^+K^-) +{\rm Br}(B_d^0\to \pi^-K^+) },
\nonumber
}
\dis{
&A^{\rm CP}_{\OVER{B}_s^{\,0}\to K^+\pi^-}=\frac{{\rm Br}(\OVER{B}_s^{\,0}\to K^+\pi^-) -{\rm Br}(B_s^0\to K^-\pi^+)}{{\rm Br}(\OVER{B}_s^{\,0}\to K^+\pi^-) +{\rm Br}(B_s^0\to K^-\pi^+) }.\nonumber
}

\begin{figure}[!t]
  \begin{center}
  \begin{tabular}{c}
   \includegraphics[width=0.3\textwidth]{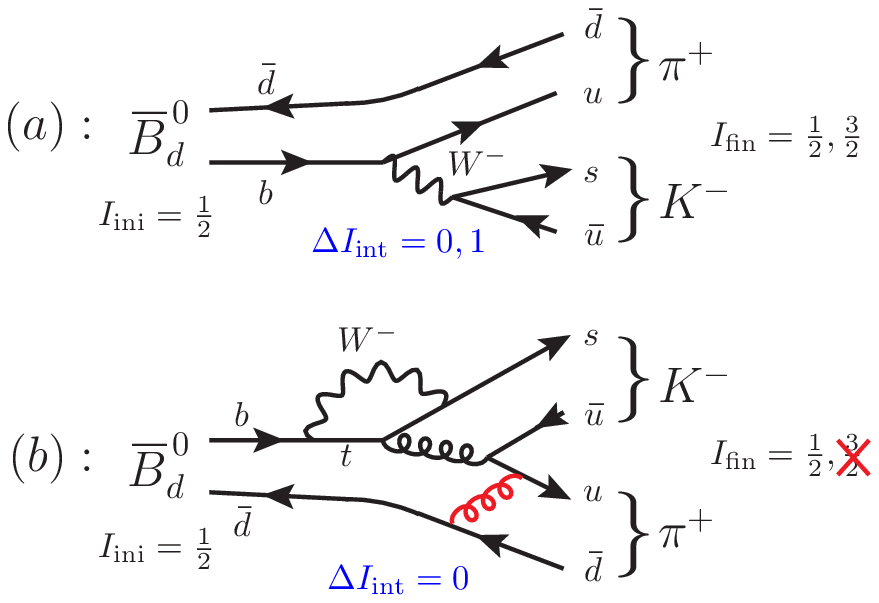}\\ \\
     \includegraphics[width=0.3\textwidth]{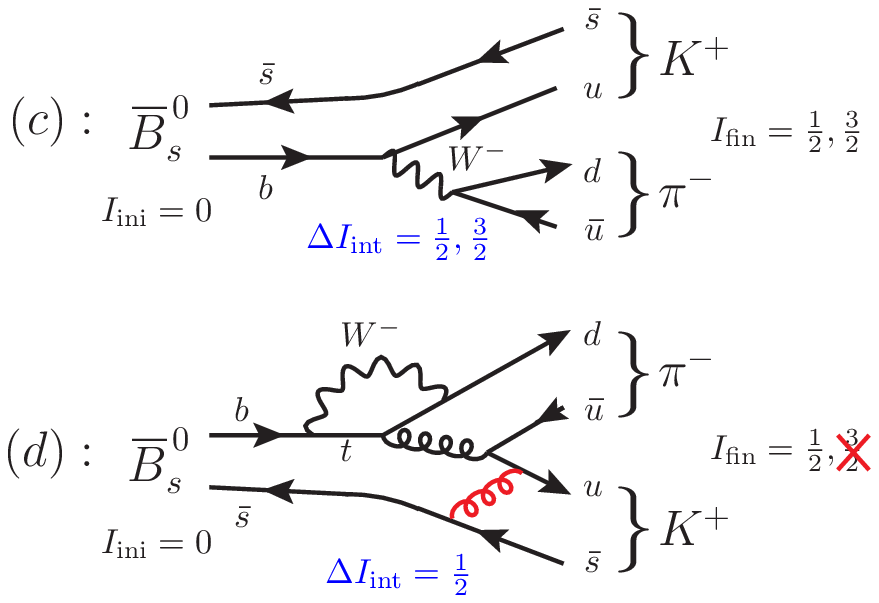}
   \end{tabular}
  \end{center}
 \caption{Diagrams contributing to the direct CP violation, p. 203 of Ref. \cite{BigiSandaBk}. With the KS parametrization, the CKM factors are, $(a)\, K_a, (b)\, K_b , (c)\, K_c$, and  $(d)\, K_d$. For (b) and (d), the final states satisfy $I_{\rm fin}=\frac12$.
  }\label{fig:FSIinclBd}
\end{figure}

For the $\OVER{B}_d^{\,0}$ decay to $\pi^+K^-$, the tree diagram plus the penguin diagram are shown in Fig. \ref{fig:FSIinclBd}. The CKM parameter dependence of the $W$-boson exchange diagrams,  (a) and (c) of Fig. \ref{fig:FSIinclBd}, are  $K_a=s_1^2c_3s_3 \simeq 0.793\times 10^{-3}$, and  $  K_c =c_1s_1s_3  \simeq 3.426 \times 10^{-3}$, respectively.  The CKM parameter dependence of the penguin diagrams, (b) and (d) of Fig. \ref{fig:FSIinclBd}, are $ K_b \simeq -c_2^2c_3s_3  + c_1c_2 s_2c_3^2 e^{-i\delta}-(15.56 +38.93\,i) \times 10^{-3}$, and $K_d \simeq -c_2c_3s_1s_2 e^{-i\delta}\simeq (9.008\,i) \times 10^{-3}$, respectively.

The interaction Hamiltonian of the $W$-exchange diagrams, \ie (a,c)  of Figs. \ref{fig:FSIinclBd}, are   $\Delta I_{\rm int}=0,1$ and $\Delta I_{\rm int}=\frac12, \frac32$, respectively. The penguin diagrams, \ie (b,d)  of Figs. \ref{fig:FSIinclBd}, are $\Delta I_{\rm int}=0$ and $\Delta I_{\rm int}=\frac12$, respectively. For the final states, $I_{\rm fin}=\frac12$ is allowed for all four cases and $I_{\rm fin}=\frac32$ is allowed only for (a) and (c). As shown with the red gluons in (b) and (d) of Fig. \ref{fig:FSIinclBd}, the colors recombine to make $K\pi$, which is an O($\alpha_s$) effect.
As the first approximation, we will neglect the electroweak penguin contributions to the FSI phases. In addition, note that the FSI phases of (a) and (c) are negligible compared to those of (b) and (d) because the FSI in (a) and (c) should involve more than one gluon, and hence
our main concern is the FSI phases from (b) and (d) \cite{Deshpande95, Buras92}. In Refs.  \cite{Deshpande95, Buras92} there are detailed studies on these with $6\times 6$ matrix, which will be used below. In addition by comparing the upper two figures and the lower two figures of Figs. \ref{fig:FSIinclBd}, we note that the final state interaction phases $\FSI{i}$ of (a,b) and (c,d) of Fig. \ref{fig:FSIinclBd} are opposite.

The strong amplitudes $A^{(s)}_{\rm w, pg}$ consist of the products of Wilson coefficients and matrix elements of the operators. The strong amplitude $A^{(s)}_{\rm w}$ can be split into two, one $\Delta I= \frac12 $ and the other $\Delta I= \frac32 $, both of which have the chiral structure $(V-A)\cdot(V-A)$.  On the other hand, the amplitude $A^{(s)}_{\rm pg}$ contains only $\Delta I= \frac12 $ piece only and the chiral structure is $(V-A)\cdot(V\mp A)$. We define $A_{\rm w}\equiv A_{\rm w}^{(1/2)}e^{i\FSI{1/2}}+ A_{\rm w}^{(3/2)}e^{i\FSI{3/2}}, A_{\rm pg}\equiv  A_{\rm pg}^{(1/2)}e^{i\FSI{\rm pg}}, A^s_{\rm w}\equiv A_{\rm w}^{s(1/2)}e^{-i\FSI{1/2}}+ A_{\rm w}^{s(3/2)}e^{-i\FSI{3/2}}$ and $A^s_{\rm pg}\equiv  A_{\rm pg}^{s(1/2)}e^{-i\FSI{\rm pg}}$, and let $A_{\rm w}^{(1/2,3/2)}, A_{\rm pg}^{(1/2)}, A_{\rm w}^{s(1/2,3/2)}$ and $A_{\rm pg}^{s(1/2)}$ be real.
Note that we commented above that $ \FSI{s,1/2}=-\FSI{1/2} ,  \FSI{s,3/2}= -\FSI{3/2}$ and $  \FSI{s,\rm pg}=-\FSI{\rm pg}$. Thus, the decay amplitudes for  $\OVER{B}_d^{\,0}\to K^+\pi^-$ and $B_d^0\to K^-\pi^+$   have the expression, using Eq. (\ref{eq:KSexact}),
$T(\OVER{B}_d^{\,0}\to K^+\pi^-) \simeq A_{\rm w} c_3 s_1^2s_3 + A_{\rm pg}( -c_2^2c_3s_3+c_1c_2c_3^2 s_2 e^{-i\delta})$,
$T(B_d^0\to \pi^+K^-)  \simeq A_{\rm w} c_3 s_1^2s_3 + A_{\rm pg}( -c_2^2c_3s_3+c_1c_2 c_3^2s_2 e^{i\delta}) ,
T(\OVER{B}_s^{\,0}\to K^+\pi^-) \simeq A_{\rm w}^s c_1 s_1s_3 + A_{\rm pg}^s(- c_2 c_3s_1 s_2 e^{-i\delta}) $, and $
T(B_s^0\to \pi^+K^-) \simeq A_{\rm w}^s c_1 s_1s_3 + A_{\rm pg}^s(- c_2 c_3s_1 s_2 e^{i\delta})$.
Since the Jarlskog invariant phase $\delta$ has been determined rather accurately in this paper, we can obtain a relation between three FSI phases, $ \FSI{1/2}, \FSI{3/2} $ and $ \FSI{\rm pn} $.
The $\Delta I=\frac12$ rule in the $B_{d,s}^0$ decay is not expected to be as strong as the $\Delta I=\frac12$ rule in the $K^0$ decay since the energies of the outgoing light mesons are much above the strong interaction scale. For the $B_{d,s}^0$ decay at the LHC energy, we use $A_{\rm w}^{(3/2)}=\xi A_{\rm w}^{(1/2)}$ (Eq. (\ref{eq:xi})) and $A_{\rm w}^{s(3/2)}=\eta A_{\rm w}^{s(1/2)}$ (Eq. (\ref{eq:eta}))  where $\xi^{-1}$ and $\eta^{-1}$
will be in the range O(1) which will appear as the results of running  the Wilson coefficient and taking the hadronic matrix elements.

Thus, the amplitudes for $\OVER{B}_d^{\,0}$ decays are,
\dis{
T&(\OVER{B}_d^{\,0}\to \pi^+K^-)\simeq  c_3 s_1^2s_3  \big(\frac{A_{\rm w}^{(1/2)}(1+\xi\cos\tilde\phi)}{\cos\tilde{\phi} }\big)\\
  &\cdot e^{i(\FSI{1/2}+ \tilde{\phi} )} +  A_{\rm pg}^{(1/2)}e^{i\FSI{\rm pg}}( -c_2^2c_3s_3+c_1c_2c_3^2 s_2 e^{-i\delta})\\
T&(B_d^0\to K^+\pi^-)\simeq c_3 s_1^2s_3  \big(\frac{A_{\rm w}^{(1/2)}(1+\xi\cos\tilde\phi)}{\cos \tilde{\phi}}\big) \\
  &\cdot e^{i(\FSI{1/2}+ \tilde{\phi} )} +  A_{\rm pg}^{(1/2)}e^{i\FSI{\rm pg}}( -c_2^2c_3s_3+c_1c_2c_3^2 s_2 e^{i\delta}).\nonumber
}
Similarly, the $\OVER{B}_s^{\,0}$ decay amplitudes are calculated. Let us define
\dis{
\tan\tilde{\phi}  &=  \xi  \,\frac{-\sin(\FSI{1/2}-\FSI{3/2})}{{1+ \xi  \cos (\FSI{3/2}-\FSI{1/2})}},\\[0.3em]
\tan\tilde{\phi}_{ s } &=  -\eta \,\frac{-\sin(\FSI{1/2}-\FSI{3/2})}{{1+ \eta \cos (\FSI{3/2}-\FSI{1/2})}}.\label{eq:tanphitilde}
}
Then, the CP asymmetries are
\dis{
A^{\rm CP}_{\OVER{B}_{[s]}^{\,0}\to \underline{K^+\pi^-}[\pi^-K^+]}=A^{\rm CP}_{\OVER{B}_{[s]}^{\,0}  }\equiv \frac{N_{[s]}}{D_{[s]}},
}
where $[~~]$ replaces the underlined part for the case of [$s$], and
\dis{
N_{[s]}&=2{\cal A}_{[s]} A_{\rm pg}^{[s](1/2)}\,\underline{c_1c_2c_3 s_2}\,[c_2c_3s_2]\,\sin\delta\\
&\cdot\sin(\tilde\phi_{[s]}+\FSI{1/2}-\FSI{\rm pg}),\\
D_{[s]}&=  2{\cal A}_{[s]}  A_{\rm pg}^{[s](1/2)}\,\underline{(c_1c_2c_3 s_2\cos\delta-c_2^2s_3)} \,[-c_2c_3s_2\cos\delta]\\
& \cdot\cos(\tilde\phi_{[s]}+\FSI{1/2}-\FSI{\rm pg})+\left({\cal A}_{[s]} \right)^2 +\left(A_{\rm pg}^{[s](1/2)}\right)^2\\
 &\cdot \underline{(c_1^2 c_2^2c_3^2s_2^2 +c_2^4s_3^2-2c_1c_2^3c_3s_2s_3\cos\delta)}\,[c_2^2c_3^2s_2^2],
 \nonumber
}
 which are functions of the {\bf w} asymmetries ${\cal A}_{[s]}$ and the {\bf pg} asymmetry  $A_{\rm pg}^{[s](1/2)}$,
\begin{widetext}
\dis{
{\cal A} = A_{\rm w}^{(1/2)}\frac{1+  {\xi} \, \cos\tilde{\phi} }{\cos\tilde{\phi} }\,
 {s_1^2s_3} ,~{\cal A}_{s}= A_{\rm w}^{s(1/2)}\frac{1+  \eta \cos\tilde{\phi}_{s}}{\cos\tilde{\phi}_{s}}\,
 c_1s_3 \,. \label{eq:Acal}
}

\begin{figure}[!t]
  \begin{center}
  \begin{tabular}{c}
  \includegraphics[width=0.95\textwidth]{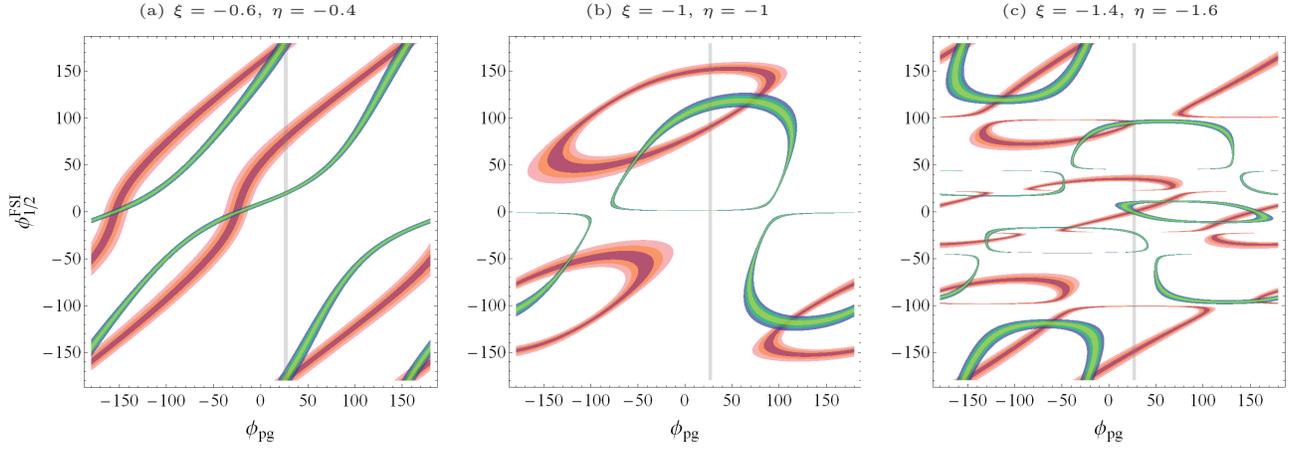}
  \end{tabular}
  \end{center}
\caption{The allowed regions in the  $\FSI{1/2}$  vs. $\FSI{\rm pg}$ plane for $r_d=r_s=26$ for (a) $\xi=-0.6,\,\eta=-0.4$, (b) $\xi=-1,\,\eta=-1$, and (c) $\xi=-1.4,\,\eta=-1.6$.
The green($1\sigma$)--blue($3\sigma$) tones are for $A^{\rm CP}_{\OVER{B}^{\,0}}$ and the brown($1 \sigma$)--lavender($3 \sigma$) tones are for $A^{\rm CP}_{\OVER{B}_{s}^{\,0}}$.
The FSI calculation is shown as the vertical gray bar for $\FSI{\rm pg} \simeq 27^0$.
  }
\label{fig:FSIBd}
\end{figure}

We determine $\theta_{1}= (13.025_{-0.038}^{+0.039})^{0}$ from $V_{11}^{\rm PDG}$ \cite{PDG12}. Then, from  $V_{21}^{\rm PDG}$ and $V_{13}^{\rm PDG}$, we determine $\theta_2=(2.292^{+2.625}_{-2.292})^{0}$ and $\theta_3=(0.8923^{+0.0382}_{-0.0357})^{0}$, respectively, but $V_{21}^{\rm PDG}$ is not very effective in determining $\theta_2$. So, from $V_{21}^{\rm PDG}, V_{31}^{\rm PDG}$ and $J^{\rm PDG}=(2.96^{+0.20}_{-0.16})\times 10^{-5}$ \cite{PDG12}, we perform  a combined fit for $\theta_2$ and $\delta$, using $J=| c_1c_2c_3s_1^2s_2s_3\sin\delta|$, and obtain $\theta_2=4.7824^0$ and $\delta= 90.00^0$. We used these first quadrant values $\theta_{1,2,3}$ and $\delta$ in Fig. \ref{fig:FSIBd}.

\end{widetext}

To estimate the FSI phases, we use the definition of the interaction given in Eqs. (6.31, 6.32) of \cite{Deshpande95, Buras92} which is $G_F/\sqrt2$ times
\dis{
{\cal H}\propto & \,K_{\bf w}\sum_{q=d,s}\left[{\cal C}_1 Q_{1,q}^u  +{\cal C}_2 Q_{2,q}^u \right] -K_{\bf pg} \sum_{i=3}^6{\cal C}_i  Q_i .
}
where $K_{\bf w,pg}$ are the CKM angles and
\dis{
&K_{\bf w}\left\{\begin{array}{l}
Q_{1,q}^u= (\overline{b}_\alpha u_\beta)_{V-A} (\OVER{u}_\beta q_\alpha)_{V-A},\,[q=s({\rm a}),d({\rm c})],\\
Q_{2,q}^u=(\overline{b}_\alpha u_\alpha)_{V-A} (\OVER{u}_\beta q_\beta)_{V-A},\,[q=s({\rm a}),d({\rm c})],
\end{array}\right.\\
&K_{\bf pg}\left\{\begin{array}{l}
Q_{3,q}=(\overline{b}_\alpha q_\alpha)_{V-A} (\OVER{u}_\beta u_\beta)_{V-A},\,[q=s({\rm b}),d({\rm d})],\\
Q_{4,q}=(\overline{b}_\alpha q_\beta)_{V-A} (\OVER{u}_\beta u_\alpha)_{V-A},\,[q=s({\rm b}),d({\rm d})],\\
Q_{5,q}=(\overline{b}_\alpha q_\alpha)_{V-A} (\OVER{u}_\beta u_\beta)_{V+A},\,[q=s({\rm b}),d({\rm d})],\\
Q_{6,q}=(\overline{b}_\alpha q_\beta)_{V-A} (\OVER{u}_\beta u_\alpha)_{V+A},\,[q=s({\rm b}),d({\rm d})],\end{array}\right.\nonumber 
}
where $q=s$\,[Fig. \ref{fig:FSIinclBd}\,(a)] and $q=d$\,[Fig. \ref{fig:FSIinclBd}\,(c)] for $Q_{1,q}^u$ and $Q_{2,q}^u$, and $q=s$\,[Fig.  \ref{fig:FSIinclBd}\,(b)] and $q=d$\,[Fig. \ref{fig:FSIinclBd}\,(d)] for $Q_i\,(i=3,\cdots,6)$. The tree level Wilson coefficients ${\cal C}_i$ have been calculated in  \cite{Deshpande95}, giving ${\cal C}_i=-0.3125, 1.1502, 0.0174, -0.0373, 0.0104, -0.0459$ for $i=1,\cdots, 6$, respectively. The one loop correction  for $\Lambda_{\overline{\rm MS}}^{(5)}=223\,$MeV corrects these by ${\cal C}_1'={\cal C}_1, {\cal C}_2'={\cal C}_2, {\cal C}_3'={\cal C}_3 -P_s/3=-0.00302+0.00712\, i, {\cal C}_4'={\cal C}_4+P_s=0.02095-0.02136\, i, {\cal C}_5'={\cal C}_5-P_s/3=-0.01002+0.00712\, i, {\cal C}_6'={\cal C}_6+P_s=0.00535-0.02136\, i$, where $P_s(5\gev)\simeq 0.0102[\frac{10}{9}+G(m_c, \mu, q^2)]\simeq 0.06127-0.02136\, i$ \cite{PsCalc}.

For the isospin study of $\overline{B}^{\,0}$ decay , consider Fig.  \ref{fig:FSIinclBd}\,(a),
\dis{
T_{a}^{\rm w} \equiv    {\bf I}_{\rm 1/2} (1+\xi),~\xi=\frac{A_{\rm w}^{(3/2)}}{A_{\rm w}^{(1/2)}} =\frac{\sqrt2\,R_{3/2}}{3\,{\bf I}_{\rm 1/2}},\label{eq:xi}
}
where $T_{a}^{\rm w} $ is $\langle K^-,\pi^+ | (\Delta I^0_{\rm int})+(\Delta I^1_{\rm int})_0 |\overline{B}_d^{\,0}\rangle$, ${\bf I}_{\rm 1/2} =\sqrt{ {2}/{3}}\, ( I_0
-           I^{1/2}_1 )$,
$I_0 = \langle I_{\frac12} | (\Delta I^0_{\rm int}) |\overline{B}_d^{\,0}\rangle,
I_1^{1/2}   \equiv \sqrt{ {1}/{3}}\,R_{1/2}$, and
$I_1^{3/2}  \equiv \sqrt{ {2}/{3}}\,R_{3/2}.$
Note also that
$| K^-,\pi^+ \rangle =
-\sqrt{ {2}/{3}}\,|I_\frac12\rangle +\sqrt{ {1}/{3}}\,|I_{\frac32}\rangle$ and $(\Delta I_{\rm int}^1)_0 |\overline{B}_d^{\,0}\rangle = \sqrt{ {2}/{3}}\,|I_{\frac32}\rangle+ \sqrt{ {1}/{3}}\,|I_{\frac12}\rangle.$

For the isospin study of $\overline{B}_s^{\,0}$ decay , consider Fig.  \ref{fig:FSIinclBd}\,(c),
\dis{
T_{c}^{\rm w}\equiv & \langle K^+,\pi^- | (\Delta I^{1/2}_{\rm int})_{1/2} +(\Delta I^{3/2}_{\rm int})_{1/2}  |\overline{B}_s^{\,0}\rangle\\
&  = {I}^{\rm 1/2} (1+\eta),
~ \eta=\frac{A_{\rm w}^{s(3/2)}}{A_{\rm w}^{s(1/2)}} =\frac{I^{3/2}}{\sqrt3\, I^{1/2}},\label{eq:eta}
}
where the reduced matrix elements are
$I^{1/2} =\sqrt{ {2}/{3}}\,\langle I_{\frac12} | (\Delta I^{1/2}_{\rm int}) |\overline{B}_s^{\,0}\rangle$ and $I^{3/2}=\langle I_{\frac32} | (\Delta I^{3/2}_{\rm int}) |\overline{B}_s^{\,0}\rangle.$

Taking the matrix elements of the $\overline{B}_d^{\,0} \to K^-\pi^+$ decay,
\dis{
A&_{\overline{B}_d^{\,0}}^{\rm CP} \propto   \frac{1}{27} [(-0.313)_{\bf 1}K_a +(1.150)_{\bf 2} K_a\\
&+(-0.003 +0.007 \, i)_{\bf 3} K_b +(0.021+0.021\, i)_{\bf 4} K_b\\
 &+(-0.010 +0.007 \, i)_{\bf 5} K_b+(0.021-0.021 \, i)_{\bf 6}  K_b ]\\
& \propto 26.15_{\bf w}K_a +(e^{26.6^{\,0}  \,i})_{\bf pg}K_b . \label{eq:BdRunning}
}
On the other hand,
\dis{
A_{\overline{B}_d^{\,0}}^{\rm CP} \to &K^-\pi^+) \propto \left[ {\bf I}_{\rm 1/2}(1+\xi)K_a +A_{\rm pg} e^{i\,\phi_{\rm pg}}K_b \right]\\
&\propto r_d (1+\xi )K_a +e^{i\,\phi_{\rm pg}}K_b,~~r_d\equiv \frac{ {\bf I}_{\rm 1/2}}{ A_{\rm pg} } .\label{eq:rd}
}
Comparing (\ref{eq:BdRunning}) and (\ref{eq:rd}), we estimate
\dis{
r_d (1+\xi )\simeq 26.15,~~ \phi_{\rm pg}  \simeq 26.6^{\,0}.
}

For the $\overline{B}_s^{\,0} \to K^+\pi^-$ decay, similarly we have
\dis{
A_{\overline{B}_s^{\,0}}^{\rm CP} \propto r_s (1+\eta )K_c +e^{i\,\phi_{\rm pg}}K_d,~ r_s = \frac{I^{1/2}}{A_{\rm pg}},\label{eq:rs}
}
and $r_s (1+\eta)\simeq 26.15,\, \phi_{\rm pg}  \simeq 26.6^{\,0}$.

In Fig. \ref{fig:FSIBd}, we present the allowed regions of $A^{\rm CP}_{\OVER{B}^{\,0}_{d,s}}$
in the  $\FSI{1/2}$  vs. $\FSI{\rm pg}$ plane for $r_d=r_s=26$. $\{\xi,\eta\}$ are (a) $\{-0.6,-0.4\}$,  (b) $\{-1,-1\}$, and  (c) $\{-1.4,-1.6\}$.
The green($1\sigma$)--blue($3\sigma$) tones are for $A^{\rm CP}_{\OVER{B}^{\,0}}$ and the brown($1 \sigma$)--lavender($3 \sigma$) tones are for $A^{\rm CP}_{\OVER{B}_{s}^{\,0}}$.
The FSI calculation is shown as the vertical gray bar at $\FSI{\rm pg} \simeq 27^0$.
Note that $\xi$ and $\eta$ are expected to be smaller than 1. In Fig. \ref{fig:FSIBd}\,(a), a common region of  $A^{\rm CP}_{\OVER{B}^{\,0}_{d}}$ and $A^{\rm CP}_{\OVER{B}^{\,0}_{s}}$ exists at $\FSI{1/2}\approx -180^0$ and $\FSI{\rm pg}\approx 27^0$. We scanned a wide range of parameters, among which we found the common regions near $\FSI{\rm pg}\approx 27^0$ for fixed $r_{d,s}$ in the vicinity of the parameters of (a) $\FSI{1/2}\approx -180^0$ and (c) $\FSI{1/2}\approx 0^0$, which imply that the SM has reasonable solutions. A more complete analysis will be presented elsewhere.

{\it Conclusion} --
With the KS parametrization of the CKM matrix, the Jarlskog determinant is simply expressed as  $J=|{\rm Im\,}V_{31}V_{22}V_{13}|$, and $\delta$ turns out to be maximal, $\delta\simeq \frac{\pi}{2}$. From this accurate determination of the CKM phase $\delta$, the new LHCb CP asymmetries from the ${\OVER{B}_{[s]}^{\,0}\to K\pi}$ decays allow us to obtain the strong FSI phases in the direct CP violation processes.

{We thank Howard Baer and Yannis Semertzidis for helpful comments. J.E.K. thanks George Zoupanos for the hospitality offered to him at the 13th Hellenic School and Workshop (Corfu Island, Greece, 1--10 Sep. 2013) where this paper was initiated.
 J.E.K. is supported in part by the National Research Foundation (NRF) grant funded by the Korean Government (MEST) (No. 2005-0093841) and by the IBS(IBS CA1310).}


\begin{thebibliography}{99}
\bibitem{PMNScomm} On the other hand,
we find that the reduction factor of $J$ from $J_{\rm max}$ in the leptonic case is O(100) times larger than that in the quark case, which is due to the large off-diagonal terms in the PMNS matrix.
\end{thebibliography}
\end{document}